\input harvmac
\newcount\figno
\figno=0
\def\fig#1#2#3{
\par\begingroup\parindent=0pt\leftskip=1cm\rightskip=1cm\parindent=0pt
\global\advance\figno by 1
\midinsert
\epsfxsize=#3
\centerline{\epsfbox{#2}}
\vskip 12pt
{\bf Fig. \the\figno:} #1\par
\endinsert\endgroup\par
}
\def\figlabel#1{\xdef#1{\the\figno}}
\def\encadremath#1{\vbox{\hrule\hbox{\vrule\kern8pt\vbox{\kern8pt
\hbox{$\displaystyle #1$}\kern8pt}
\kern8pt\vrule}\hrule}}

\overfullrule=0pt

%
\def\underarrow#1{\vbox{\ialign{##\crcr$\hfil\displaystyle
 {#1}\hfil$\crcr\noalign{\kern1pt\nointerlineskip}$\longrightarrow$\crcr}}}
%
\def\tilde{\widetilde}

\def\inbar{\vrule height1.5ex width.4pt depth0pt}
\def\IC{\relax\hbox{\kern.25em$\inbar\kern-.3em{\rm C}$}}
\def\IR{\relax\hbox{\kern.25em$\inbar\kern-.3em{\rm R}$}}
\def\IZ{\relax\ifmmode\hbox{Z\kern-.4em Z}\else{Z\kern-.4em Z}\fi}

\font\zfont = cmss10 

\def\bigone{\hbox{1\kern -.23em {\rm l}}}
\def\ZZ{\hbox{\zfont Z\kern-.4emZ}}


\def\drawbox#1#2{\hrule height#2pt
        \hbox{\vrule width#2pt height#1pt \kern#1pt
              \vrule width#2pt}
              \hrule height#2pt}

\def\Fund#1#2{\vcenter{\vbox{\drawbox{#1}{#2}}}}
\def\Asym#1#2{\vcenter{\vbox{\drawbox{#1}{#2}
              \kern-#2pt       
              \drawbox{#1}{#2}}}}

\def\fund{\Fund{6.5}{0.4}}

\batchmode
  \font\bbbfont=msbm10
\errorstopmode
\newif\ifamsf\amsftrue
\ifx\bbbfont\nullfont
  \amsffalse
\fi
\ifamsf
\def\IR{\hbox{\bbbfont R}}
\def\IC{\hbox{\bbbfont C}}

\def\IZ{\hbox{\bbbfont Z}}


\midinsert
\endinsert


\nref\gsw{ M.B. Green, J.H. Schwarz and E. Witten, {\it Superstring Theory},
Two volumes, Cambridge Univesity Press, Cambridge (1986).}

\nref\pol{ J. Polchinski, {\it String Theory}, Two volumes, Cambridge University
Press, Cambridge (1998).} 

\nref\hatfield{B. Hatfield, {\it Quantum Field Theory of Point Particles
and Stings}, Addison-Wesley Publishing, Redwood City, (1992).}

\nref\theisen{ D. L\"ust and S. Theisen, {\it Lectures on  String Theory},
Lecture Notes in Physics {bf 346}, Springer-Verlag, Berlin (1989).}

\nref\yau{S.T. Yau (ed.), {\it Mathematical Aspects of String Theory},
World Scientific, Singapore (1987).}

\nref\greene{ S.T. Yau (ed.), {\it Mirror Symmetry I}, American Mathematical Society,
Providence USA (1998); B. Greene and S.T. Yau (eds.), {\it Mirror Symmetry II},
American Mathematical Society, Providence USA (1997).}

\nref\vafa{C. Vafa, ``Geometric Physics'', hep-th/9810149.}

\nref\ktheory{E. Witten, ``D-branes and K-Theory'', JHEP {\bf 812:019} (1998);
hep-th/9810188.}

\nref\cds{A. Connes, M. Douglas and A. Schwarz, `Noncommutative Geometry and 
Matrix Theory: Compactification on Tori'', JHEP {\bf 9802:003}  (1998), hep-th/9711162.}

\nref\schomerus{ V. Schomerus, ``D-branes and Deformation Quantization'', 
JHEP {\bf 9906:030} (1999).}

\nref\sw{N. Seiberg and E. Witten, ``String Theory and Noncommutative
Geometry'', JHEP {\bf 9909:03} (1999), hep-th/9908142.}

\nref\dito{J. Dito, Lett. Math. Phys. {\bf 20} (1990) 125; Lett. Math.
Phys. {\bf 27} (1993) 73.}

\nref\antonu{ F. Antonsen, Phys. Rev. D {\bf 56}, 920 (1997).}

\nref\antond{F. Antonsen, ``Deformation Quantization of Constrained
Systems'', gr-qc/9710021.}

\nref\antont{F. Antonsen,  ``Deformation Quantization of Gravity'',
gr-qc/9712012.}

\nref\curtright{T. Curtright D. Fairlie and C. Zachos, Phys. Rev. D 
{\bf 58},  025002 (1998).}

\nref\zachos{ T. Curtright and C. Zachos, J. Phys. A {\bf 32},
771 (1999).}

\nref\dq{H. Garc\'{\i}a-Compe\'an, J.F. Pleba\'nski, M. Przanowski and
F.J. Turrubiates, ``Deformation Quantization of Classical Fields'',
hep-th/9909206.}

\nref\pita{ V.B. Berestetskii, E.M. Lifshitz and L.P. Pitaevskii,
{\it Relativistic Quantum Theory}, (Pergamon Press, New York, 1971). }

\nref\weyl{H. Weyl, {\it Group Theory and Quantum Mechanics}, (Dover, New
York, 1931).}

\nref\wigner{E.P. Wigner, Phys. Rev. {\bf 40}, 749 (1932).}

\nref\moyal{J.E. Moyal, Proc. Camb. Phil. Soc. {\bf 45}, 99 (1949).}

\nref\strato{R.L. Stratonovich, Sov. Phys. JETP {\bf 31} (1956) 1012.}

\nref\aw{G.S. Agarwal and E. Wolf, Phys. Rev. D {\bf 2} (1970) 2161;
2206.}

\nref\bayen{F. Bayen, M. Flato, C. Fronsdal, A. Lichnerowicz and D.
Sternheimer, Ann. Phys. {\bf 111}, 61 (1978); Ann. Phys. {\bf 111}, 111
(1978).}

\nref\tata{W.I. Tatarskii, Usp. Fiz. Nauk {\bf 139} (1983) 587.}

\nref\hillery{M. Hillery, R.F. O'Connell, M.O. Scully and E.P. Wigner,
Phys. Rep. {\bf 106}, 121 (1984).}

\nref\carinena{J.F. Cari\~nena, J.M. Gracia Bond\'{\i}a and J.C. Varilly,
{\it J. Phys. A: Math. Gen.} {\bf 23}, 901 (1990).}

\nref\gadella{M. Gadella, Fortschr. Phys. {\bf 43} (1995) 229.}

\nref\ppt{ J.F. Pleba\'nski, M. Przanowski and J. Tosiek, {\it Acta Phys. Pol.}
{\bf B27} 1961 (1996).}

\nref\bordemann{ M. Bordemann, N. Neumaier and S. Waldman, J. Geom. Phys. {\bf 29}
(1999) 199.}

\nref\key{B.S. Kay, {\it Phys. Rev.} {\bf D20}, 3052 (1979).}

\nref\birrel{ N.D. Birrell and P.C.W. Davis, {\it Quantum Fields in Curved Space}
(Cambridge University Press, Cambridge, 1982).}

\nref\isham{C.J. Isham, {\it Proc. R. Soc. Lond.} {\bf A362}, 383 (1978).}

\nref\fradkin{ I.A. Batalin and E.S. Fradkin, Ann. Inst. Henri Poincar\'e,
{\bf 49}, 145 (1988).}

\nref\ferra{ S. Ferrara and M.A. Lled\'o, ``Some Aspects of Deformation of Supersymmetric
Field Theories'', hep-th/0002084.}

\nref\tera{ S. Terashima, ``A Note on Superfields and Noncommutative Geometry'', 
hep-th/0002119.}



\Title{hep-th/0002212, CINVESTAV-FIS-00/08}
{\vbox{\centerline{}
\medskip
\centerline{Deformation Quantization of Bosonic Strings}}}
\smallskip
\centerline{H.
Garc\'{\i}a-Compe\'an,$^{a}$\foot{compean@fis.cinvestav.mx} J.F. 
Pleba\'nski,$^a$\foot{pleban@fis.cinvestav.mx} M.
Przanowski$^{b,a}$\foot{przan@fis.cinvestav.mx} and
F.J. Turrubiates$^a$\foot{fturrub@fis.cinvestav.mx} }
\smallskip
\centerline{\it $^a$Departamento de F\'{\i}sica}
\centerline{\it  Centro de Investigaci\'on y de Estudios Avanzados del
IPN}
\centerline{\it Apdo. Postal 14-740, 07000, M\'exico D.F., M\'exico}
\smallskip
\centerline{\it $^b$Institute of Physics}
\centerline{\it Technical University of \L \'od\'z}
\centerline{\it W\'olcza\'nska 219, 93-005, \L \'od\'z, Poland}
\bigskip
\baselineskip 18pt
\medskip
\vskip 1truecm
\noindent
Deformation quantization of bosonic strings is considered. We show that 
the light-cone gauge is the most convenient classical description to perform 
the quantization of 
bosonic strings in the deformation quantization formalism. Similar to the 
field theory case, the oscillator variables greatly facilitates the analysis. 
The mass spectrum, propagators and the Virasoro algebra are finally described 
within this deformation quantization scheme.


\noindent

\Date{February, 2000}

\newsec{Introduction} 

String theory is one of the most beautiful attempts to reconcile quantum mechanics and
general relativity (for a review see \refs{\gsw,\pol,\hatfield,\theisen}). From the physical
point of view it is our best understanding of all matter and their interactions into an
unified scheme. On the mathematical side, string theory has been used to motivate
unsuspected interplay among some mathematical subjects. At the perturbative level, it is
well known that string theory is related to the theory of Riemann surfaces \yau\ and some
aspects of algebraic geometry and Mirror symmetry (see for instance \greene).

Non-perturbative revolution of string theory, through the introduction of D-branes and
duality, have shown to be related to some aspects of toric geometry (see for instance
\vafa, K-theory (see the Witten seminal paper, \ktheory), noncommutative geometry and
deformation quantization theory \refs{\cds,\schomerus,\sw}. This latter is relevant for
the description of the low energy effective theory of open strings on the D-brane
world-volume, when a non-zero Neveu-Schwarz $B$-field is introduced. Thus, in this
context, deformation quantization describes properly the noncommutative space time
instead of the standard quantization of the phase space of the
two-dimensional conformal field theory.

However, it would be important to describe the noncommutativity of space time and the
quantization within the same framework of deformation quantization theory. In order to do
that one has to quantize first string theory within this framework. 
This is something which has not been done yet in the literature. And at the same
time this is the natural extension of the deformation quantization of classical fields
developed in [12-18].

The purpose of the present paper is to give the first step into this unified description of
quantum and noncommutativity in string theory. We will consider here the deformation
quantization of the bosonic string theory. The fermionic case is leave for a forthcoming
paper. In Sec. 2 we overview the preliminaries and notation of string theory in order to
prepare the theory for quantization. Sec. 3 is devoted to do the quantization of bosonic
strings by using deformation quantization theory. In Sec. 4 we describe the Casimir effect
and the normal ordering within deformation quantization formalism. Correlation functions and
Green functions for the bosonic string are computed in Sec. 5. Finally in Sec. 6 we give our
final comments.

\vskip 2truecm

\newsec{Classical Strings}

In this section we give a brief review of the classical bosonic string theory
(for further details see \refs{\gsw,\pol,\hatfield,\theisen}). 

We consider the string world-sheet $\Sigma$ embedded into the $D$-dimensional
space-time $M$ of Lorentzian metric $\eta_{\mu \nu} = diag(-1,1,\dots ,1),$
$\mu, \nu=0,1, \dots ,D-1$. This embedding is defined by

\eqn\uno{
X^{\mu} = X^{\mu} (\sigma^a), \ \ \ \ \  \mu = 0,1, \dots , D-1 \ \ {\rm and}
\ \ a =0,1 }
where $X^{\mu}$ are the space-time coordinates and $\sigma^a$ stand for the 
coordinates on the world-sheet $\Sigma$. (In what follows the Greek indices 
$\mu, \nu, \dots = 0,1, \dots , D-1$ correspond to the space-time components, the
Latin indices $a,b, \dots = 0,1$ refer to the world-sheet coordinates). 

Let $g_{ab}$ be a
Riemannian metric of the Lorentzian signature $(-,+)$ on $\Sigma$. To find the 
equations of motion for the string we use the Polyakov action

\eqn\dos{
S_P =S_P[X^{\mu}, g_{ab}]= - {T \over 2} \int_{\Sigma} d^2 \sigma
 \sqrt{-g} g^{ab} \partial_a
X^{\mu} \partial_b X_{\mu}, }
where $T$ denotes the string tension. Then the equations of motion read 

\eqn\tres{
 {\delta S_P \over \delta X^{\mu}}=0 \Longleftrightarrow \partial_a
\big( \sqrt{-g} g^{ab} \partial_b X^{\mu} \big) = 0}
and

\eqn\cuatro{
{\delta S_P \over \delta g^{ab}}=0 \Longleftrightarrow 
T_{ab}:= - {2 \over \sqrt{-g}}{\delta S_P \over \delta g^{ab}} 
=  T \bigg(\partial_a X^{\mu} \partial_b X_{\mu} 
- {1 \over 2} g_{ab}  g^{cd} \partial_c X^{\mu} \partial_d X_{\mu}\bigg) = 0.}
Of course, Eqs. \tres\ are the Laplace-Beltrami equations for $X^{\mu}$ on 
$\Sigma$, and Eqs. \cuatro, saying that the energy-momentum tensor $T_{ab}$
vanishes, are the constraints of the string theory. One can quickly find that
the constraint equations \cuatro\ give

\eqn\cinco{
 g_{ab} = \Omega^2 \partial_a X^{\mu} \partial_b X_{\mu}}
with $\Omega = \Omega(\sigma^a)$ being an arbitrary nowhere vanishing 
function on $\Sigma$. The relation \cinco\ means that the metric $g_{ab}$ is 
conformally equivalent to the induced metric $h_{ab} = \partial_a X^{\mu}
\partial_b X_{\mu}$ on $\Sigma$. The form of the action \dos\ tells us that the
theory is invariant with respect to: $(i)$ Poincar\'e transformations on $M$,
$(ii)$ diffeomorphisms of $\Sigma$ and $(iii)$ conformal rescalings of $g_{ab}$.
The $(ii)$ and $(iii)$ symmetries  enable one to choose the coordinates 
$\sigma^a$ and the function  $\Omega$ so that (conformal gauge)

$$ (\sigma^0, \sigma^1) \equiv (\tau, \sigma), \ \ \ \  0 \leq \sigma 
\leq \pi, $$

\eqn\seis{
 (g_{ab}) = \pmatrix{-1 & 0 \cr
0 & 1\cr}. }
With \seis\ assumed the equations of motion \tres\ take the simple form of
the Klein-Gordon equations

\eqn\siete{
 \ddot{X}^{\mu} - X''^{\mu}  = 0, }
where $\dot{X}^{\mu} \equiv \partial_{\tau} X^{\mu}$ and $X'^{\mu} \equiv
\partial_{\sigma}X^{\mu}.$

Then the constraint equations \cuatro\ read 

$$ T_{01} = T_{10} = 0 \Longleftrightarrow \dot{X}^{\mu} \dot{X}_{\mu} = 0$$

\eqn\ocho{
T_{00} = T_{11} = 0 \Longleftrightarrow \dot{X}^{\mu} \dot{X}_{\mu} +
X'^{\mu} X'_{\mu}= 0.}
The Polyakov action \dos\ is now

\eqn\nueve{
S_P = - {T \over 2} \int d\tau \int_0^{\pi} d \sigma \bigg( - \dot{X}^{\mu}
\dot{X}_{\mu} + X'^{\mu} X'_{\mu} \bigg). }
In the case of the {\it open string, i.e.} $X^{\mu}(\tau,0) \not= 
X^{\mu}(\tau,\pi)$ the action \nueve\ leads to the equations \siete\ if the 
following boundary conditions are imposed for each $\tau$

\eqn\diez{
 X'^{\mu}(\tau,0) = 0, \ \ \ \ \ {\rm and} \ \ \ \ \ X'^{\mu}(\tau,\pi) =0.}
For the {\it closed string} one has from the very definition 

\eqn\once{
 X^{\mu}(\tau,0) = X^{\mu}(\tau,\pi).}
We will deal separately with the closed and open strings.

\vskip 1truecm
\subsec{Closed Strings}

The general solution of Eqs. \siete\ satisfying the boundary conditions \once\
can be written in the form of the following series

\eqn\doce{
 X^{\mu}(\tau, \sigma) = x^{\mu} + {1 \over  \pi T}p^{\mu}\tau
+ {1 \over \sqrt{2 \pi T}} \sum_{n \not=0} \sqrt{{\hbar \over 2|n|}}
 \bigg\{ a^{\mu}_n
exp\bigg( 2i(n \sigma - |n| \tau) \bigg) +  a^{\mu *}_n
exp\bigg( -2i (n \sigma - |n|\tau) \bigg) \bigg\}, } 
where $x^{\mu}$ and $p^{\mu}$ are real and the star $``*''$ stands for the 
complex conjugation.

The conjugate momentum $\Pi^{\mu}$ of $X^{\mu}$ is as usual defined  by

$$
\Pi^{\mu}(\tau,\sigma) = \eta^{\mu \nu} {\delta L \over \delta 
\dot{X}^{\mu}(\tau,\sigma)} = T \dot{X}^{\mu}(\tau,\sigma) 
$$
\eqn\trece{
=
{1 \over \pi} p^{\mu} + i \sqrt{ T \over 2 \pi} \sum_{n \not= 0} 
\sqrt{2 \hbar |n|} \bigg\{ a^{\mu *}_n
exp\bigg( -2i(n \sigma - |n| \tau) \bigg) -  a^{\mu}_n
exp\bigg( 2i (n \sigma - |n|\tau) \bigg) \bigg\},}
where $L=L[X^{\mu},\dot{X}^{\mu}]$ is the Lagrangian 

\eqn\catorce{
 L = L[X^{\mu}, \dot{X}^{\mu}]= - {T \over 2}\int_0^{\pi} 
d \sigma \bigg( - \dot{X}^{\mu}
\dot{X}_{\mu} + X'^{\mu} X'_{\mu} \bigg). }
Then the standard Poisson brackets

$$\{ X^{\mu}(\tau,\sigma), \Pi^{\nu}(\tau,\sigma ') \} 
= \eta^{\mu \nu}\delta 
(\sigma - \sigma '), $$

\eqn\quince{
 \{ X^{\mu}(\tau, \sigma), X^{\nu} (\tau, \sigma ') \} = 0 = 
\{ \Pi^{\mu} (\tau, \sigma), \Pi^{\nu}(\tau,\sigma ') \} }
lead to the following Poisson brackets for $x^{\mu}, \ p^{\mu},\ a^{\mu}_n$
and  $a^{\mu *}_n$

\eqn\dseis{
\{x^{\mu}, p^{\nu} \} = \eta^{\mu \nu}, \ \ \ \ \{a^{\mu}_m, a^{\nu *}_n \}
= - {i \over \hbar} \delta_{mn} \eta^{\mu \nu} }
with the remaining independent Poisson brackets being zero.

From Eqs. (2.12) and (2.13) we get

\eqn\dsiete{
x^{\mu} = {1 \over \pi} \int_0^{\pi} d \sigma \ X^{\mu}(0,\sigma) \ \ \ \ {\rm and} \ \ \ \ 
 p^{\mu} = \int_0^{\pi} d \sigma \ \Pi^{\mu}(\tau, \sigma).}
Thus $x^{\mu}$ and $p^{\mu}$ are the the center-of-mass position at $\tau=0$ and momentum 
of the string. Consequently

\eqn\docho{
M^2 = - p^{\mu} p_{\mu}}
is the square mass of the string.

In string theory it is convenient to use instead of $a^{\mu}_n$ and $a^{\mu *}_n$
the following objects

$$ \alpha^{\mu}_n = -i \sqrt{\hbar n} \ a^{\mu}_n, \ \ \ \ \ \ \ 
\tilde{\alpha}^{\mu}_n = -i \sqrt{\hbar n} \ a^{\mu}_{-n}, $$

\eqn\dnueve{
 \alpha^{\mu}_{-n} = \alpha^{\mu *}_n = i \sqrt{\hbar n} \ a^{\mu *}_n, \ \ \ \ \ \ \
\tilde{\alpha}^{\mu}_{-n} = \tilde{\alpha}^{\mu *}_n= i \sqrt{\hbar n} \ a^{\mu *}_{-n}}
for $n > 0$. Substituting Eq. \dnueve\ into Eqs. \doce, \trece\ and \dseis\ one has

\eqn\veinte{
 X^{\mu}(\tau, \sigma) = x^{\mu} + {1 \over  \pi T}p^{\mu}\tau
+ {i \over 2\sqrt{\pi T}} \sum_{n \not=0} {1 \over n}
 \bigg\{ \alpha^{\mu}_n
exp\bigg( - 2in (\tau - \sigma) \bigg) +  \tilde{\alpha}^{\mu}_n
exp\bigg( -2i n ( \tau + \sigma) \bigg) \bigg\},} 

\eqn\vuno{
 \Pi^{\mu}(\tau,\sigma) = 
{1 \over \pi} p^{\mu} +  \sqrt{ T \over \pi} \sum_{n \not= 0} 
\bigg\{ \alpha^{\mu}_n
exp\bigg( -2in(\tau - \sigma) \bigg) +  \tilde{\alpha}^{\mu}_n
exp\bigg( - 2in (\tau +\sigma) \bigg) \bigg\}}
and

$$\{x^{\mu}, p^{\nu} \} = \eta^{\mu \nu},$$

\eqn\vdos{
\{\alpha^{\mu}_m, \alpha^{\nu}_n \}= - im \delta_{m+n,0} \eta^{\mu \nu},\ \ \ \ \ \ 
\{\tilde{\alpha}^{\mu}_m, \tilde{\alpha}^{\nu}_n \}= - im \delta_{m+n,0} 
\eta^{\mu \nu}}
for all $m,n \not= 0$.

We now consider the so called {\it light-cone gauge} in which the constraint
equations \ocho\ can be easily solved and then eliminated. This gauge will be crucial
for the deformation quantization of the bosonic string.

First, introduce the light-cone (null) coordinates 

\eqn\vtres{
  X^{\pm} := {1 \over \sqrt{2}}( X^0 \pm X^{D-1}),}
and the remaining coordinates $X^j,$ $j=1, \dots , D-2$ are left as before. As
$X^+(\tau,\sigma)$ satisfies the wave equation \siete\ one can choose the 
coordinate $\tau$ in such a manner that

\eqn\vcuatro{
X^+(\tau,\sigma) = {1 \over \pi T} p^+ \tau }
and \seis\ still holds true. In this gauge the constraint equations \ocho\ are
equivalent to the following equations

\eqn\vcinco{
(\dot{X}^{\mu} - X'^{\mu})(\dot{X}_{\mu} - X'_{\mu}) =0 \Longleftrightarrow
\sum_{j=1}^{D-2} (\dot{X}^j - X'^{j})^2 - {2 p^+ \over \pi T}
(\dot{X}^- - X'^{-}) = 0,}          

\eqn\vseis{
(\dot{X}^{\mu} + X'^{\mu})(\dot{X}_{\mu} + X'_{\mu}) =0 \Longleftrightarrow 
\sum_{j=1}^{D-2} (\dot{X}^j + X'^{j})^2 - {2 p^+ \over \pi T}
(\dot{X}^- + X'^{-}) = 0.}
Expressing $X^-(\tau,\sigma)$ in the form 

\eqn\vsiete{
X^-(\tau,\sigma) = x^- + {1 \over  \pi T}p^{-}\tau
+ {i \over 2\sqrt{\pi T}} \sum_{n \not=0} {1 \over n}
 \bigg\{ \alpha^{-}_n
exp\bigg( - 2in (\tau - \sigma) \bigg) +  \tilde{\alpha}^{-}_n
exp\bigg( -2i n ( \tau + \sigma) \bigg) \bigg\},} 
and inserting \vsiete\ into \vcinco\ and \vseis\ one can solve these constraint
equations in the sense that $p^-$, $\alpha^-_n$ and $\tilde{\alpha}^-_n$ are
defined by $p^+$, $p^j,$ $\alpha^j_n$ and $\tilde{\alpha}^j_n$. Simple 
calculations give 

\eqn\vocho{
\alpha^-_m = {\sqrt{\pi T} \over p^+} \sum_{j=1}^{D-2} \sum_{n=-\infty}^{\infty}
\alpha^j_n \alpha^j_{m-n}}
and

\eqn\vnueve{
\tilde{\alpha}^-_m = {\sqrt{\pi T} \over p^+} \sum_{j=1}^{D-2} \sum_{n=-\infty}^{\infty}
\tilde{\alpha}^j_n \tilde{\alpha}^j_{m-n},}
for all $m \in {\IZ}$ and  where we denote 

\eqn\treinta{
\alpha^-_0 = \tilde{\alpha}^-_0 := {1 \over 2 \sqrt{\pi T}} p^- \ \ \ \ \ \   {\rm and}
\ \ \ \ \ \ \alpha^j_0 = \tilde{\alpha}^j_0 := {1 \over 2 \sqrt{\pi T}} p^j.}
Thus the (independent) dynamical variables of the string are: $x^-$, $p^+$, $x^j$, $p^j$, 
$\alpha^j_n$ and $\tilde{\alpha}^j_n$ for $n \not= 0$ or, equivalently: $x^-$, $p^+$,
$X^j$ and $\Pi^j$.

For the Poisson bracket for $x^-$ and $p^+$ we have 

\eqn\tuno{
\{x^-,p^+\} = -1.}
(Note the opposite sign in \tuno\ in comparison with the usual one, $\{x^j,p^k\} = 
\delta^{jk}$).

The square mass $M^2$ given by \docho\ takes now the form 

$$
M^2 = 2 p^+p^- - \sum_{j=1}^{D-2} p^jp^j = 4 \pi T \sum_{j=1}^{D-2} \sum_{n= - \infty}^{\infty}
\alpha^j_n \alpha^j_{-n} - 4 \pi T \sum_{j=1}^{D-2} \alpha^j_0 \alpha^j_0
$$

\eqn\tdos{ = 4 \pi T \sum_{j=1}^{D-2} \sum_{n \not= 0}
\alpha^j_n \alpha^j_{-n}
= 4 \pi T \sum_{j=1}^{D-2} \sum_{n \not= 0}
\tilde{\alpha}^j_n \tilde{\alpha}^j_{-n}.}
Then the Hamiltonian

\eqn\ttres{
H = {T \over 2} \int_0^{\pi} d\sigma \bigg \{ \sum_{j=1}^{D-2} \bigg(({\Pi^{j} \over T})^2 +
(X'^{j})^2\bigg) \bigg\} }
is

$$
H = \sum_{j=1}^{D-2} \sum_{n = - \infty}^{\infty} \bigg( \alpha^j_n \alpha^j_{-n}
+ \tilde{\alpha}^j_n \tilde{\alpha}^j_{-n} \bigg)
$$

$$ = 2 \sum_{j=1}^{D-2} \sum_{n = - \infty}^{\infty} \alpha^j_n \alpha^j_{-n} 
$$

$$ = { \sum_{j=1}^{D-2} (p^j)^2 \over 2 \pi T } + 4 \sum_{j=1}^{D-2} \sum_{n=1}^{\infty}
\alpha^j_n \alpha^j_{-n}
$$
\eqn\tcuatro{  = { \sum_{j=1}^{D-2} (p^j)^2 \over 2 \pi T } + 2 \hbar \sum_{j=1}^{D-2}
\sum_{n \not= 0} |n| a^{j*}_n a^j_n = {p^+ p^- \over \pi T}. }

From Eqs. \vocho, \vnueve\ and \vdos\ one gets
$$
\{ \alpha^-_m, \alpha^-_n \} = -i {2 \sqrt{\pi T} \over p^+} (m-n) \alpha^-_{m+n},
$$

\eqn\tcinco{
\{ \tilde{\alpha}^-_m, \tilde{\alpha}^-_n \} = -i {2 \sqrt{\pi T} \over p^+} (m-n) 
\tilde{\alpha}^-_{m+n}}
for all $m,n \not= 0$. Therefore $\{ \alpha^-_m; \ {\rm for \ all} \ m \not= 0\}$ and  
$\{ \tilde{\alpha}^-_m; \ {\rm for \ all} \ m \not= 0\}$ constitute  the {\it Virasoro algebra
without a central extension.} Now analogously as in the case of classical fields [19,18]
we introduce the oscillator variables $Q^j_n$ and $P^j_n$, $n \not= 0,$ as follows

$$
Q^j_n(\tau):= \sqrt{ \hbar \over 4 |n|} \bigg( a^j_n(\tau) + a^{j*}_n(\tau) \bigg)
$$

\eqn\tseis{
P^j_n(\tau) := i \sqrt{ \hbar|n|} \bigg( a^{j*}_n(\tau) - a^j_n(\tau) \bigg),}
where $a^j_n(\tau) := a^j_n exp \big( -2i|n| \tau \big).$

By Eq. \dseis\ one has 

$$
\{Q^j_m(\tau), P^k_n(\tau) \} = \delta^{jk} \delta_{mn},
$$

\eqn\tsiete{ \{Q^j_m(\tau), Q^k_n(\tau)\} = 0 = \{P^j_m(\tau), P^k_n(\tau)\}.}

From Eq. \tseis\ we quickly find

\eqn\tocho{
a^j_n(\tau) = \sqrt{|n| \over \hbar} \bigg( Q^j_n(\tau) + {i \over 2 |n|} P^j_n(\tau) \bigg).}
Straightforward calculations show that Eqs. \doce\ and \trece\ give

\eqn\tnueve{
a_n^j(\tau) = {1 \over 2 \sqrt{\pi \hbar |n|}} \int_0^{\pi} d \sigma
\bigg( 2 |n| \sqrt{T} X^j(\tau,\sigma) + {i \over \sqrt{T}} \Pi^j(\tau,\sigma) \bigg)
exp \bigg( -2in \sigma \bigg). }
Substituting Eq. \tnueve\ into Eq. \tseis\ one gets

$$
Q^j_n(\tau) = {1 \over \sqrt{\pi}} \int_0^{\pi} d \sigma \bigg( \sqrt{T} X^j(\sigma)
\cos (2n \sigma + 2 |n| \tau) + { 1\over 2 |n| \sqrt{T}} \Pi^j(\sigma) \sin 
(2n \sigma + 2 |n| \tau) \bigg),
$$

\eqn\cuarenta{
P^j_n(\tau) = {1 \over \sqrt{\pi}} \int_0^{\pi} d \sigma \bigg( -2 |n| \sqrt{T}
 X^j(\sigma)
\sin (2n \sigma + 2 |n| \tau) + { 1\over \sqrt{T}} \Pi^j(\sigma) \cos  
(2n \sigma + 2 |n| \tau) \bigg),}
where $X^j(\sigma) \equiv X^j(0,\sigma)$ and $\Pi^j(\sigma) \equiv \Pi^j(0,\sigma).$

Inserting Eq. \tocho\ into \doce\ and \trece\ we obtain

$$
X^j(\tau,\sigma) = x^j + {1\over \pi T} p^j \tau  + {1 \over \sqrt{\pi T}} 
\sum_{n \not= 0} \bigg( Q^j_n \cos (2n \sigma - 2 |n| \tau) - 
{1 \over 2 |n|}P^j_n \sin (2n \sigma - 2 |n| \tau) \bigg),
$$

\eqn\cuno{
\Pi^j(\tau,\sigma) = {1\over \pi} p^j   + \sqrt{T \over \pi } 
\sum_{n \not= 0} \bigg( 2 |n| Q^j_n \sin (2n \sigma - 2 |n| \tau) + 
P^j_n \cos (2n \sigma - 2 |n| \tau) \bigg), }
where $Q^j_n \equiv Q^j_n(0)$ and $P^j_n \equiv P^j_n(0).$

Observe also that from Eq. \tseis\ one quickly finds that

$$
Q^j_n(\tau) = Q^j_n \cos (2|n| \tau) + {1 \over 2 |n|} P^j_n \sin (2|n| \tau),
$$
\eqn\cdos{
P^j_n(\tau) = -2 |n| Q^j_n \sin (2|n| \tau) + P^j_n \cos (2|n| \tau).}
Finally, Eqs. \tdos\ and \tcuatro\ with \tocho\ give 

\eqn\ctres{
M^2 = 4\pi T \sum_{j=1}^{D-2} \sum_{n \not=0} \bigg( (P^j_n)^2 + 4 n^2 
(Q^j_n)^2 \bigg)}
and

\eqn\ccuatro{  
H = { \sum_{j=1}^{D-2} (p^j)^2 \over 2 \pi T } + {1 \over 2} \sum_{j=1}^{D-2}
\sum_{n \not= 0} \bigg( (P^j_n)^2 + 4 n^2
(Q^j_n)^2 \bigg).}
We arrive at the conclusion that one can use the (independent) dynamical variables
$(x^-,p^+,x^j,p^j,Q^j_n,P^j_n)$ and these variables are canonically related to the
variables $(x^-,p^+,X^j, \Pi^j).$ 

\noindent
[Straightforward calculations give 

$$ \{X^j(\tau,\sigma),\Pi^k(\tau,\sigma ')\}_{(x,p,Q,P)} : = \sum_{l=1}^{D-2}
\bigg\{\bigg( 
{\partial X^j(\tau,\sigma) \over \partial x^l}
{\partial \Pi^k(\tau,\sigma ') \over \partial p^l} -
{\partial X^j(\tau,\sigma) \over \partial p^l}
{\partial \Pi^k(\tau,\sigma ') \over  \partial x^l}\bigg) +
$$
$$
\sum_{n \not= 0} \bigg(
{\partial X^j(\tau,\sigma) \over \partial Q^l_n}
{\partial \Pi^k(\tau,\sigma ' ) \over \partial P^l_n} -
{\partial X^j(\tau,\sigma) \over \partial P^l_n}
{\partial \Pi^k(\tau,\sigma ') \over  \partial Q^l_n}\bigg)\bigg\}
$$

$$
= {1\over \pi} \delta^{jk} + \delta^{jk} \bigg( \delta(\sigma - \sigma ') - {1\over \pi}\bigg)
= \delta^{jk} \delta(\sigma - \sigma '), 
$$

\eqn\ccinco{ 
\{X^j(\tau,\sigma),X^k(\tau,\sigma ')\}_{(x,p,Q,P)}=0=
\{\Pi^j(\tau,\sigma),\Pi^k(\tau,\sigma ')\}_{(x,p,Q,P)}.]}

\vskip 1truecm
\subsec{Open Strings}

In this case the general solution of Eqs. \siete\ satisfying the boundary conditions
\diez\ can be represented by the series

$$
 X^{\mu}(\tau, \sigma) = x^{\mu} + {1 \over  \pi T}p^{\mu}\tau
+ {1 \over \sqrt{ \pi T}} \sum_{n=1}^{\infty} \sqrt{{\hbar \over n}}
 \bigg( a^{\mu}_n
exp\big( -in \tau \big) +  a^{\mu *}_n
exp\big( i n \tau \big) \bigg) \cos(n \sigma) 
$$ 

\eqn\cseis{
= x^{\mu} + {1 \over  \pi T}p^{\mu}\tau
+ {i \over \sqrt{ \pi T}} \sum_{n \not= 0} {1 \over n}
 \alpha^{\mu}_n
exp\big( -in \tau \big) \cos (n \sigma).}
Note that the boundary condition \diez\ at $\sigma = 0$ yields $a^{\mu}_n = a^{\mu}_{-n}$
for all $n \not= 0.$ Here $\alpha^{\mu}_n$ are defined as before by \dnueve\ and
$\tilde{\alpha}^{\mu}_n$ do not appear as independent variables because $a^{\mu}_n =
a^{\mu}_{-n}$. 

Then 

$$
\Pi^{\mu}(\tau,\sigma) = T \dot{X}^{\mu} =
{1 \over \pi} p^{\mu} + i \sqrt{ T \over  \pi} \sum_{n=1}^{\infty} 
\sqrt{ \hbar n} \bigg( a^{\mu *}_n
exp\big( in \tau \big) -  a^{\mu}_n
exp\big( -in \tau \big) \bigg) \cos (n \sigma)
$$ 

\eqn\csiete{
={1 \over \pi} p^{\mu} + \sqrt{ T \over  \pi} \sum_{n \not= 0} 
\alpha^{\mu}_n
exp\big( -in \tau \big) \cos (n \sigma).}
In the light-cone gauge we have 

$$ X^+ = {1 \over \pi T} p^+ \tau$$

\eqn\cocho{
X^- = x^{-} + {1 \over  \pi T}p^{-}\tau
+ {i \over \sqrt{ \pi T}} \sum_{n \not= 0} {1 \over n}
 \alpha^{\mu}_n
exp\big( -in \tau \big) \cos (n \sigma).}
Then the solution of the constraint equations \ocho\ or, equivalently, \vcinco\
and \vseis\ reads

\eqn\cnueve{
\alpha^-_m = {\sqrt{\pi T} \over 2p^+} \sum_{j=1}^{D-2} \sum_{n=-\infty}^{\infty}
\alpha^j_n \alpha^j_{m-n}}
for all $m \in \IZ$, where now

\eqn\cincuenta{
\alpha^-_0 := {1 \over  \sqrt{\pi T}} p^- \ \ \ \ \ \   {\rm and}
\ \ \ \ \ \ \alpha^j_0 := {1 \over  \sqrt{\pi T}} p^j.}
(Compare with Eq. \treinta.)

The square mass $M^2$ is 

\eqn\ciuno{
M^2 = - p^{\mu}p_{\mu} = 2 p^+ p_- - \sum_{j=1}^{D-2} p^jp^j = \pi T \sum_{j=1}^{D-2}
\sum_{n \not= 0}
\alpha^j_n \alpha^j_{-n}}
and the Hamiltonian \ttres\ reads

$$ H = {1 \over 2} \sum_{j=1}^{D-2} \sum_{n=-\infty}^{\infty}
\alpha^j_n \alpha^j_{-n}
$$

\eqn\cidos{  = { \sum_{j=1}^{D-2} (p^j)^2 \over 2 \pi T } +  \hbar \sum_{j=1}^{D-2}
\sum_{n=1}^{\infty} n a^{j}_n a^{j*}_n.}

For the Poisson brackets $\{\alpha^-_m, \alpha^-_n \}$ one gets

\eqn\citres{
\{ \alpha^-_m, \alpha^-_n \} = -i { \sqrt{\pi T} \over p^+} (m-n) \alpha^-_{m+n}}
for all $m,n \not= 0.$ (Compare with \tcinco).

Analogously as before we introduce new variables $Q^j_n$ and $P^j_n$, $n=1, \dots , \infty$

$$
Q^j_n(\tau):= \sqrt{ \hbar \over 2 n} \bigg( a^j_n(\tau) + a^{j*}_n(\tau) \bigg)
= {i \over n \sqrt{2}} \bigg( \alpha^j_n(\tau) - \alpha^j_{-n}(\tau)\bigg)$$

\eqn\cicuatro{
P^j_n(\tau) := i \sqrt{ \hbar n \over 2} \bigg( a^{j*}_n(\tau) - a^j_n(\tau) \bigg)
={1 \over \sqrt{2}} \bigg( \alpha^j_n(\tau) + \alpha^j_{-n}(\tau)\bigg),}
where $a^j_n(\tau) := a^j_n exp \big( -in \tau \big),$ $n \in\IZ_+.$
$Q^j_n(\tau)$ and $P^j_n(\tau)$ fulfill the Poisson bracket formulas (2.37). Then

\eqn\cicinco{
a^j_n(\tau) = \sqrt{n \over 2 \hbar} \bigg( Q^j_n(\tau) + {i \over n} P^j_n(\tau) \bigg),
\ \ \ \  n \in \IZ_+.}

Inserting Eq. (2.55) into Eqs. (2.46) and (2.47) one gets

$$
X^j(\tau,\sigma) = x^j + {1 \over \pi T} p^j \tau + \sqrt{2 \over \pi T}
\sum_{n=1}^{\infty} \bigg( Q^j_n \cos (n \tau) + {1 \over n} P^j_n \sin (n \tau)
\bigg) \cos (n \sigma),
$$

\eqn\ciseis{
\Pi^j(\tau,\sigma) = {1 \over \pi } p^j + \sqrt{2T \over \pi }
\sum_{n=1}^{\infty} \bigg(-n Q^j_n \sin (n \tau) +  P^j_n \cos (n \tau)
\bigg) \cos (n \sigma),}
where, as before, $Q^j_n \equiv Q^j_n(0)$ and $P^j_n \equiv P^j_n(0).$

From Eqs. (2.46) and (2.47) we find

\eqn\cisiete{
a^j_n(\tau) = {1 \over \sqrt{\pi \hbar n}} \int_0^{\pi} d \sigma \ 
\bigg( n \sqrt{T} X^j(\tau,\sigma) + {i \over \sqrt{T}} \Pi^j(\tau,\sigma) \bigg) \cos (n
\sigma).}
Substituting Eq. (2.57) into (2.54) one quickly obtains 

$$ 
Q^j_n(\tau) = \sqrt{2 \over \pi} \int_0^{\pi} d\sigma \  \bigg( \sqrt{T} X^j(\sigma)
\cos (n \tau) + {1 \over n \sqrt{T}} \Pi^j(\sigma) \sin (n\tau)\bigg) \cos (n \sigma),
$$

\eqn\ciocho{
P^j_n(\tau) = \sqrt{2 \over \pi} \int_0^{\pi} d\sigma \  \bigg( - n \sqrt{T} X^j(\sigma)
\sin (n \tau) + {1 \over \sqrt{T}} \Pi^j(\sigma) \cos (n\tau)\bigg) \cos (n \sigma).}

From Eq. (2.58) we have 

$$ 
Q^j_n(\tau) = Q^j_n \cos (n \tau) + {1 \over n} P^j_n \sin (n \tau),
$$

\eqn\cinueve{
P^j_n(\tau) = -n Q^j_n \sin (n \tau) +  P^j_n \cos (n \tau),}
(compare with Eq. (2.42)).

Finally, for $M^2$ and $H$ one gets

\eqn\sesenta{
M^2 = \pi T \sum_{j=1}^{D-2} \sum_{n=1}^{\infty} \bigg( (P^j_n)^2 + n^2 (Q^j_n)^2 \bigg)}
and

\eqn\suno{
H = { \sum_{j=1}^{D-2} (p^j)^2 \over 2 \pi T} + {1 \over 2} \sum_{j=1}^{D-2}
\sum_{n=1}^{\infty} \bigg( (P^j_n)^2 + n^2 (Q^j_n)^2 \bigg).}
As before we can use the (independent) dynamical variables $(x^-,p^+,x^j,p^j,Q^j_n,P^j_n)$
and they are canonically related to $(x^-,p^+,X^j,\Pi^j).$ Observe that in the present
case $n \in \IZ_+.$

\vskip 2truecm

\newsec{Deformation Quantization of the Bosonic String}

In this section we are going to use the well known machinery of deformation 
quantization [20-31] to the case of bosonic strings.

\vskip 1truecm
\subsec{Closed Strings}

According to results of subsection $2.1$ the phase space ${\cal Z}$ of a closed 
string can be understood as the Cartesian product ${\cal Z} = \IR^2 \times
\IR^{2(D-2)} \times \IR^{2 \infty}$ endowed with the following symplectic form 

\eqn\sdos{
\omega = d p_- \wedge dx^- + \sum_{j=1}^{D-2} \bigg( dp_j \wedge dx^j +
\sum_{n \not= 0} dP_{jn} \wedge dQ^j_n \bigg),}
where $p_- = -p^+,$ $p_j = p^j$ and $P_{jn} = P^j_n.$

Equivalently, one can consider ${\cal Z}$ to be ${\cal Z} = \IR^2 \times \Gamma$
where $\Gamma$ is the set $\Gamma = \{ \big( X^j(\sigma), \Pi_j(\sigma) \big)_{j=1,\dots ,
D-2} \}$
with $X^j(\sigma)$ and $\Pi^j(\sigma)= \Pi_j(\sigma)$ being arbitrary real functions of
$\sigma \in [0,\pi]$ satisfying the boundary conditions: $X^j(0) = X^j(\pi)$ and $\Pi^j(0)=
\Pi^j(\pi)$. The symplectic form  has now the functional form

\eqn\stres{
\omega = dp_- \wedge  dx^- + \sum_{j=1}^{D-2} \int_0^{\pi} d \sigma \ \delta \Pi_j (\sigma)
\wedge \delta X^j(\sigma).}
Let $\hat{x}^-,$ $\hat{p}^+ = - \hat{p}_-$, $\hat{X}^j$ and $\hat{\Pi}^j$ be the field
operators

$$ 
\hat{x}^- |x^- \rangle = x^- |x^- \rangle , \ \ \ \ \ \ \hat{p}^+ |p^+ \rangle 
= p^+ |p^+ \rangle ,
$$

$$ 
\hat{X}^j(\sigma) |X^j \rangle = X^j(\sigma) |X^j \rangle , \ \ \ \ \ \ \hat{\Pi}^j(\sigma)
|\Pi^j\rangle = \Pi^j(\sigma) |\Pi^j \rangle ,
$$

\eqn\scuatro{
[\hat{X}^j(\sigma),\hat{\Pi}^k(\sigma ')] = i \hbar \delta^{jk} \delta( \sigma - \sigma '),
\ \ \ \ \ \ \ \ \  [\hat{x}^-, \hat{p}^+] = -i \hbar.}
Denote

$$
|x^-,X \rangle := |x^- \rangle \otimes \bigg( \bigotimes_{j=1}^{D-2} |X^j \rangle \bigg),
\ \ \ \ \ \ \ \ |p^+, \Pi \rangle := |p^+ \rangle \otimes \bigg( \bigotimes_{j=1}^{D-2} |\Pi^j
\rangle \bigg),
$$

\eqn\scinco{
{\cal D} X = \prod_{\sigma} d X^1(\sigma) \dots dX^{D-2}(\sigma) \ \ \ \ \ \ {\rm and}
\ \ \ \ \ \ {\cal D} \Pi = \prod_{\sigma} d \Pi^1(\sigma) \dots d \Pi^{D-2} (\sigma).}
Then we fix the normalization as follows

\eqn\sseis{
\int d x^- {\cal D} X | x^-,X \rangle \langle x^-,X| = \hat{1} \ \ \ \ \ {\rm and}
\ \ \ \ \ \int d \big({p^+ \over 2 \pi \hbar} \big) {\cal D} ({\Pi \over 2 \pi \hbar}\big)
|p^+, \Pi \rangle \langle p^+, \Pi| = \hat{1}.}

Let $F = F[x^-,X,p^+,\Pi]$ be a functional on the phase space ${\cal Z}$. Then according
to the Weyl rule we assign the following operator  $\hat{F}$ corresponding 
to $F$

\eqn\ssiete{
\hat{F} = W(F)= \int {dx^- dp^+ \over 2 \pi  \hbar} {\cal D} X {\cal D} ( {\Pi \over 2 \pi
\hbar}) F[x^-,X,p^+,\Pi] \hat{\Omega}[x^-,X,p^+,\Pi],}
where $\hat{\Omega}[x^-,X,p^+,\Pi]$ is the {\it Stratonovich-Weyl quantizer} (SW)

$$
\hat{\Omega}[x^-,X,p^+,\Pi] = \int d \xi^- {\cal D} \xi exp \bigg\{ -{i\over \hbar}
\bigg( - \xi^- p^+ + \int_0^{\pi} d \sigma \ \xi(\sigma) \cdot \Pi(\sigma) \bigg) \bigg\}
$$
$$
|x^- - {\xi^- \over 2}, X - {\xi \over 2} \rangle \langle X  + {\xi \over 2}, x^- + 
{\xi^- \over 2}|
$$

\eqn\socho{
= \int d ({\eta^+\over 2 \pi \hbar}) {\cal D} ({\eta \over 2 \pi \hbar})
exp \bigg\{ - {i \over \hbar} \bigg( - x^- \eta^+ + \int_0^{\pi} d \sigma \eta(\sigma)
\cdot X(\sigma) \bigg) \bigg\} | p^+ + {\eta^+ \over 2}, \Pi + {\eta \over 2} \rangle \langle
\Pi - {\eta \over 2}, p^+ - {\eta^+ \over 2} | }
with the obvious notation $ \xi(\sigma) \cdot \Pi(\sigma) \equiv \sum_{j=1}^{D-2} \xi^j(\sigma)
\Pi^j(\sigma)$ and  $\eta(\sigma) \cdot X(\sigma) \equiv \sum_{j=1}^{D-2} \eta^j(\sigma) X^j(\sigma).$

The SW quantizer has the properties 

\eqn\snueve{
\bigg( \hat{\Omega}[x^-,X,p^+,\Pi]\bigg)^{^{\dag}} = \hat{\Omega}[x^-,X,p^+,\Pi],}

\eqn\setenta{
Tr \bigg(\hat{\Omega}[x^-,X,p^+,\Pi]\bigg) = 1,}

\eqn\seuno{
Tr \bigg(\hat{\Omega}[x^-,X,p^+,\Pi] \hat{\Omega}[{'x}^-,{'X},{'p}^+,{'\Pi}]\bigg) =
\delta (x^- -{'x}^-) \delta ({p^+ -{'p}^+ \over 2 \pi \hbar}) \delta[X-{'X}] \delta
[{\Pi - {'\Pi} \over 2 \pi \hbar}].}
Multiplying Eq. (3.6) by $\hat{\Omega}[x^-,X,p^+,\Pi]$ and taking the trace one has

\eqn\sedos{
W^{-1}(\hat{F}) = F[x^-,X,p^+,\Pi] = Tr \bigg( \hat{\Omega}[x^-,X,p^+,\Pi] \hat{F} \bigg).}

This enables us to solve  the following problem. Let $F_1 = F_1[x^-,X,p^+,\Pi]$ and  $F_2 =
F_2[x^-,X,p^+,\Pi]$ be functionals defined  on the phase space ${\cal Z}$ and let
$\hat{F}_1= W(F_1)$ and $\hat{F}_2 =W(F_2)$ be their corresponding operators. The problem is
what a functional on ${\cal Z}$ corresponds to the product $\hat{F}_1 \hat{F}_2.$ This functional
is denoted by $F_1 * F_2$ and it is called the {\it Moyal $*$-product of} $F_1$ $and$ $F_2.$

By Eq. (3.11) one gets

\eqn\setres{
\big(F_1 * F_2 \big) [x^-,X,p^+,\Pi] := W^{-1}(\hat{F}_1 \hat{F}_2)= Tr \bigg(
\hat{\Omega}[x^-,X,p^+,\Pi]
\hat{F}_1 \hat{F}_2 \bigg).}
Substituting Eq. (3.6) into (3.12), using then (3.7) and perfoming straightforward but tedious
manipulations (see for example [30]) we finally obtain

$$\big(F_1 *  F_2\big) [x^-,X,p^+,\Pi] =  F_1[x^-,X,p^+,\Pi]
exp\bigg\{{i\hbar\over 2} \buildrel{\leftrightarrow}\over {\cal P}\bigg\}
F_2[x^-,X,p^+,\Pi],$$

\eqn\secuatro{
\buildrel{\leftrightarrow}\over {\cal P} := 
\bigg({{\buildrel{\leftarrow}\over {\partial}}\over
\partial p^{+}} {{\buildrel{\rightarrow}\over {\partial}}\over
\partial x^{-}} - {{\buildrel{\leftarrow}\over {\partial}}\over
\partial x^{-}} {{\buildrel{\rightarrow}\over {\partial}}\over
\partial p^{+}}\bigg) + \sum_{j=1}^{D-2}
\int_0^{\pi} d \sigma \ 
\bigg({{\buildrel{\leftarrow}\over {\delta}}\over
\delta X^{j}(\sigma)} {{\buildrel{\rightarrow}\over {\delta}}\over
\delta \Pi^{j}(\sigma)} - {{\buildrel{\leftarrow}\over {\delta}}\over
\delta \Pi^{j}(\sigma)} {{\buildrel{\rightarrow}\over {\delta}}\over
\delta X^{j}(\sigma)}\bigg).}

Now it is an easy matter to define the Wigner functional. Assume $\hat{\rho}$ to be
the density operator of the quantum state of a bosonic string. Then according to the
general formula (3.11) the functional $\rho[x^-,X,p^+,\Pi]$ corresponding to $\hat{\rho}$
reads (use also (3.7))

$$
{\rho} [x^-,X,p^+,\Pi]= W^{-1}(\hat{\rho})= Tr \bigg( \hat{\Omega}[x^-,X,p^+,\Pi] \hat{\rho}
\bigg)
$$

\eqn\secinco{
= \int d\xi^-  {\cal D} \xi
exp \bigg\{ -{i\over \hbar}
\bigg( - \xi^- p^+ + \int_0^{\pi} d \sigma \ \xi(\sigma)\cdot \Pi(\sigma) \bigg) \bigg\}
\langle X + {\xi \over 2}, x^- + {\xi^- \over 2}| \hat{\rho} | x^- - {\xi^- \over 2}, 
X - {\xi \over 2} \rangle .}
Then the {\it Wigner functional} ${\rho}_{_W} [x^-,X,p^+,\Pi]$ is defined by a simple
modification of Eq. (3.14).  Namely,

$$
\rho_{_W}[x^-,X,p^+,\Pi] := \int d ({\xi^- \over 2 \pi \hbar}){\cal D} ({\xi \over 2 \pi
\hbar}) exp \bigg\{ - { i \over \hbar} 
\bigg( - \xi^- p^+ +  \int_0^{\pi} d\sigma \ \xi (\sigma) \cdot
\Pi (\sigma) \bigg)\bigg\} 
$$

\eqn\seseis{
\langle X + {\xi \over 2}, x^- + {\xi^- \over 2} | 
\hat{\rho} | x^- - {\xi^- \over 2}, X - {\xi \over 2} \rangle .}
In particular for the pure state $\hat{\rho} = |\Psi\rangle \langle \Psi |$  we get

$$
\rho_{_W}[x^-,X,p^+,\Pi] = \int d ({\xi^- \over 2 \pi \hbar}){\cal D} ({\xi \over 2 \pi
\hbar})  exp \bigg\{ - { i \over \hbar}
\bigg( - \xi^- p^+ +  \int_0^{\pi} d\sigma \ \xi (\sigma)
\cdot \Pi (\sigma) \bigg)\bigg\} 
$$

\eqn\sesiete{
\Psi^*[x^- - {\xi^- \over 2}, X - {\xi \over 2}]
\Psi[x^- + {\xi^- \over 2}, X + {\xi \over 2}],}
where $\Psi[x^-,X]$ stands for $| \Psi \rangle$ in the Sch\"odinger representation.

As it will be clear very soon some calculations simplify when the variables $(x^-,p^+,
x^j,p^j,$ $Q^j_n,P^j_n)$ are used. In terms of these variables one has (in the obvious
notation)

$$
\hat{\Omega}(x^-,Q,p^+,P) = \int d \xi^-  d \xi \ 
exp \bigg\{ -{i\over \hbar}
\bigg( - \xi^- p^+ + \xi \cdot P \bigg)
\bigg\} |x^- - {\xi^- \over 2}, Q - {\xi \over 2} \rangle \langle Q  + {\xi \over 2}, x^- + 
{\xi^- \over 2}|
$$
\eqn\seocho{
= \int d ({\eta^+\over 2 \pi \hbar})  d ({\eta \over 2 \pi \hbar}) \ 
exp \bigg\{ - {i \over \hbar} \bigg( - x^- \eta^+ + \eta \cdot Q \bigg) \bigg\} 
| p^+ + {\eta^+ \over 2}, P + {\eta \over 2} \rangle \langle
P - {\eta \over 2}, p^+ - {\eta^+ \over 2} |,}
where 
$d \xi \equiv  \prod_{n \in \IZ} d \xi^1_n \dots d \xi^{D-2}_n,$
$ d({\eta \over 2 \pi \hbar}) \equiv \prod_{n \in \IZ} d ({\eta^1_n \over 2 \pi \hbar})
\dots  d ({\eta^{D-2}_n \over 2 \pi \hbar}),$
$\xi \cdot P \equiv  \sum_{j=1}^{D-2} \sum_{n = - \infty}^{\infty} \xi^j_n P^j_n,$
$ \eta \cdot Q \equiv$   $\sum_{j=1}^{D-2} \sum_{n = - \infty}^{\infty}$ $ \eta^j_n Q^j_n,$
$P^j_0 \equiv p^j$ and $Q^j_0 \equiv x^j.$

Then

$$\big(F_1 *  F_2\big) (x^-,Q,p^+,P) =  F_1(x^-,Q,p^+,P)
exp\bigg\{{i\hbar\over 2} \buildrel{\leftrightarrow}\over {\cal P}\bigg\}
F_2(x^-,Q,p^+,P),$$

\eqn\senueve{
\buildrel{\leftrightarrow}\over {\cal P} := 
\bigg({{\buildrel{\leftarrow}\over {\partial}}\over
\partial p^{+}} {{\buildrel{\rightarrow}\over {\partial}}\over
\partial x^{-}} - {{\buildrel{\leftarrow}\over {\partial}}\over
\partial x^{-}} {{\buildrel{\rightarrow}\over {\partial}}\over
\partial p^{+}}\bigg) + \sum_{j=1}^{D-2} \sum_{n= -\infty}^{\infty} 
\bigg({{\buildrel{\leftarrow}\over {\partial}}\over
\partial Q^{j}_n} {{\buildrel{\rightarrow}\over {\partial}}\over
\partial P^{j}_n} - {{\buildrel{\leftarrow}\over {\partial}}\over
\partial P^{j}_n} {{\buildrel{\rightarrow}\over {\partial}}\over
\partial Q^{j}_n}\bigg).}

We can also express the Moyal $*$-product in terms of $a^j_n$ and $a^{j*}_n$
or $\alpha^j_n$ and $\tilde{\alpha}^j_n$:

$$ * = exp\bigg\{{i\hbar\over 2} \buildrel{\leftrightarrow}\over {\cal P}\bigg\}
$$

$$ 
= exp \bigg\{{i \hbar \over 2} \bigg[\bigg({{\buildrel{\leftarrow}\over {\partial}}\over
\partial p^{+}} {{\buildrel{\rightarrow}\over {\partial}}\over
\partial x^{-}} - {{\buildrel{\leftarrow}\over {\partial}}\over
\partial x^{-}} {{\buildrel{\rightarrow}\over {\partial}}\over
\partial p^{+}}\bigg) + \sum_{j=1}^{D-2} 
\bigg({{\buildrel{\leftarrow}\over {\partial}}\over
\partial x^{j}} {{\buildrel{\rightarrow}\over {\partial}}\over
\partial p^{j}} - {{\buildrel{\leftarrow}\over {\partial}}\over
\partial p^{j}} {{\buildrel{\rightarrow}\over {\partial}}\over
\partial x^{j}}\bigg) \bigg] \bigg \}$$

$$ exp \bigg\{ {1 \over 2}
\sum_{j=1}^{D-2} \sum_{n \not= 0}
\bigg({{\buildrel{\leftarrow}\over {\partial}}\over
\partial a^{j}_n} {{\buildrel{\rightarrow}\over {\partial}}\over
\partial a^{j*}_n} - {{\buildrel{\leftarrow}\over {\partial}}\over
\partial a^{j*}_n} {{\buildrel{\rightarrow}\over {\partial}}\over
\partial a^{j}_n}\bigg)\bigg\}
$$

\eqn\ochenta{
= \dots exp \bigg\{ {\hbar \over 2}
\sum_{j=1}^{D-2} \sum_{n \not= 0}
n \bigg({{\buildrel{\leftarrow}\over {\partial}}\over
\partial \alpha^{j}_n} {{\buildrel{\rightarrow}\over {\partial}}\over
\partial \alpha^{j}_{-n}} + {{\buildrel{\leftarrow}\over {\partial}}\over
\partial \tilde{\alpha}^{j}_n} {{\buildrel{\rightarrow}\over {\partial}}\over
\partial \tilde{\alpha}^{j}_{-n}}\bigg)\bigg\}. }

Finally, for the Wigner function one obtains

$$
\rho_{_W}(x^-,Q,p^+,P) = \int d ({\xi^- \over 2 \pi \hbar}) d ({\xi \over 2 \pi
\hbar}) exp \bigg\{ - { i \over \hbar} 
\bigg( - \xi^- p^+ +  \xi \cdot P \bigg)\bigg\} 
$$

\eqn\ouno{
\langle Q + {\xi \over 2}, x^- + {\xi^- \over 2} | 
\hat{\rho} | x^- - {\xi^- \over 2}, Q - {\xi \over 2} \rangle }
and in the case of the pure state $\hat{\rho} = |\Psi\rangle \langle \Psi |$  

$$
\rho_{_W}(x^-,Q,p^+,P) = \int d ({\xi^- \over 2 \pi \hbar}) d ({\xi \over 2 \pi
\hbar})  exp \bigg\{ - { i \over \hbar}
\bigg( - \xi^- p^+ +  \xi \cdot P \bigg)\bigg\} 
$$

\eqn\odos{
\Psi^*(x^- - {\xi^- \over 2}, Q - {\xi \over 2})
\Psi(x^- + {\xi^- \over 2}, Q + {\xi \over 2}).}

Given $\rho_{_W}$ one can use Eq. (3.6) to find the coresponding density operator
$\hat{\rho}$

\eqn\otres{
\hat{\rho} = \int d x^- d p^+ d Q d P \rho_{_W}(x^-,Q,p^+,P) \hat{\Omega}(x^-,Q,p^+,P).}
Consequently, the average value $\langle \hat{F} \rangle$ reads

$$
\langle \hat{F} \rangle = { Tr \big( \hat{\rho} \hat{F} \big) \over Tr \big( \hat{\rho} \big)}
$$

$$
= { \int d x^- d p^+ dQ dP \rho_{_W}(x^-,Q,p^+,P) Tr \big( \hat{\Omega}(x^-,Q,p^+,P) \hat{F} \big)
\over \int d x^- d p^+ dQ dP \rho_{_W}(x^-,Q,p^+,P)}
$$

\eqn\ocuatro{
= { \int d x^- d p^+ dQ dP \rho_{_W}(x^-,Q,p^+,P) W^{-1} \big( \hat{F} \big)(x^-,Q,p^+,P)
\over \int d x^- d p^+ dQ dP \rho_{_W}(x^-,Q,p^+,P)}.}

Assume that $\hat{\rho} = | \Psi \rangle \langle \Psi |.$ Substituting  this $\hat{\rho}$ into Eq.
(3.22),
multiplying from the left by $\langle \tilde{Q},\tilde{x}^-|$ and from the right by $|\tilde{x}^-,
\tilde{Q} \rangle$ and 
employing Eq. (3.17) one gets

\eqn\ocinco{
| \Psi(\tilde{x}^-,\tilde{Q}) |^2 = \int d p^+ dP \rho_{_W}(\tilde{x}^-,\tilde{Q},p^+,P).}
Suppose that $\Psi(\tilde{x}^-,\tilde{Q}) \not= 0.$
Then inserting $\hat{\rho} = | \Psi \rangle \langle \Psi |$ into (3.22), multiplying
from the left by $\langle Q, x^- |$ and from the right  by $ | \tilde{x}^-, \tilde{Q} \rangle ,$ using
Eqs. (3.17) and (3.24) we easily find the wave function $\Psi(x^-,Q)$ in terms of the 
corresponding Wigner function $\rho_{_W}$

$$
\Psi(x^-,Q) = exp \bigg\{ i \varphi \bigg\}
$$
\eqn\oseis{
 {  \int d p^+ dP \rho_{_W}[{x^-+
\tilde{x}^-\over 2},{Q+
\tilde{Q}\over 2},p^+,P] 
 \ exp \bigg\{ - {i \over \hbar} \big( -(x^- - \tilde{x}^-) p^+ + (Q - \tilde{Q}) \cdot P
\big) \bigg\}
\over \bigg(\int d p^+ dP \rho_{_W}(\tilde{x}^-,\tilde{Q},p^+,P) \bigg)^{1/2}},}
where $\varphi$ is an arbitrary real constant.

Of course  in terms of $X^j(\sigma)$ and $\Pi^j(\sigma)$ one has 

$$
\Psi[x^-,X] = exp \bigg\{ i \varphi \bigg\}
$$
\eqn\osiete{
 {  \int d p^+ {\cal D}\Pi \ 
\rho_{_W}[{x^- + \tilde{x}^-\over 2},{X + \tilde{X}\over 2},p^+,\Pi] 
exp \bigg\{ - {i \over \hbar} \big( -(x^- - \tilde{x}^-) p^+ + \int_0^{\pi} d \sigma \
(X(\sigma)
- \tilde{X}(\sigma))\cdot \Pi(\sigma) \big) \bigg\}
\over \bigg(\int d p^+ {\cal D} \Pi \rho_{_W}[\tilde{x}^-,\tilde{X},p^+,\Pi] \bigg)^{1/2}},}
where $X(\sigma)\cdot  \Pi(\sigma) \equiv \sum_{j=1}^{D-2} X^j(\sigma) \Pi^j (\sigma).$

The natural question is: when a real function $\rho_{_W}(x^-,Q,p^+,P)$ repesents some
quantum state, {\it i.e.} it can be considered to be a Wigner function. The 
necessary and sufficient condition reads

\eqn\lacan{
\int dx^- dp^+ dQ dP \rho_{_W}(x^-,Q,p^+,P) \bigg( f^* * f \bigg)(x^-,Q,p^+,P) \geq 0}
for any $ f \in C^{\infty}({\cal Z})[[\hbar ]]$, and 

\eqn\lacanu{
\int dx^- dp^+ dQ dP \rho_{_W}(x^-,Q,p^+,P) > 0.}
(See [28,31]).

\vskip 1truecm
\subsec{Example: The Ground State}

The Wigner function $\rho_{_{W0}}$ of the ground state is defined by

\eqn\oocho{
a^j_n * \rho_{_{W0}} = 0, \ \ \ \ p^j * \rho_{_{W0}}=0 \ \ \ \  {\rm and} \ \ \ \ 
p^+ * \rho_{_{W0}} = 0, }
for $j=1, \dots , D-2$ and $n \not= 0.$

Employing Eq. (3.19) we have

\eqn\onueve{
a^j_n \rho_{_{W0}} + {1 \over 2} {\partial \rho_{_{W0}} \over \partial a^{j*}_n} = 0,
\ \ \  p^j \rho_{_{W0}} = 0, \ \ \  {\rm and} \ \ \  p^+ \rho_{_{W0}} = 0, }
for $j=1, \dots , D-2$ and $n \not= 0.$ The general real solution of Eq. (3.30) satisfying
also Eqs. (3.27) and (3.28) reads

\eqn\noventa{
\rho_{_{W0}} =C  exp \bigg \{ - 2  \sum_{j=1}^{D-2} \sum_{n \not= 0}
a^j_n a^{j*}_n  \bigg\} \delta (p^1) \dots \delta(p^{D-2})
\delta(p^+),}
where $C>0$. Consequently, in terms of $Q^j_n$ and $P^j_n$ one gets

\eqn\nuno{
\rho_{_{W0}} =C  exp \bigg \{ - {1 \over 2 \hbar} \sum_{j=1}^{D-2} \sum_{n \not= 0}
{1\over |n|} \bigg( (P^j_n)^2 + 4 n^2 (Q^j_n)^2 \bigg) \bigg\} \delta (p^1) \dots \delta(p^{D-2})
\delta(p^+).}
Observe that $\rho_{_{W0}}$ is defined by Eqs. (3.29), (3.27) and (3.28) uniquely up to an
arbitrary real constant factor $C > 0$.
This fact can be interpreted in deformation quantization formalism as the {\it uniqueness of the
vacuum state.}

Then any higher state can be obtained as an appropriate product of the form

$$
\big( a^{* i_1}_{n_1} \dots a^{* i_s}_{n_s} \big) *
\bigg\{ C exp \bigg( -{1 \over 2 \hbar} \sum_{j=1}^{D-2} \sum_{n \not= 0}
{1 \over |n|} \big( (P^j_n)^2 + 4 n^2 (Q^j_n)^2 \big) \bigg)
$$
\eqn\ndos{ 
\delta(p^1 - p^1_0) \dots \delta(p^{D-2} - p^{D-2}_0) \delta(p^+ - p^+_0)
\bigg\} * \big( a^{ i_s}_{n_s} \dots a^{ i_1}_{n_1} \big),}
(compare with \dq).

An interesting question is when a real function $\rho_{_W}(x^-,Q,p^+,P)$ satisfying
Eqs. (3.27) and (3.28) is the Wigner function of a pure state. The answer to this question
in the case of a system of particles can be found in a beautiful paper by
Tataskii \tata. In our case the solution is quite similar. To this end denote

\eqn\ntres{
\gamma(x^-,Q,\tilde{x}^-, \tilde{Q}):= \int d p^+ d P \rho_{_W}({x^- + \tilde{x}^- \over 2},
{Q + \tilde{Q} \over 2}, p^+,P) exp \bigg\{ {i \over \hbar}\big[- (x^- - \tilde{x}^-)p^+
+ (Q - \tilde{Q})P \big] \bigg\}.}
From Eq. (3.25) it follows that if $\rho_{_W}$ is the Wigner function of the pure state
$|\Psi \rangle \langle \Psi |$ then

$$
{\partial^2 \ln \gamma(x^-,Q,\tilde{x}^-,\tilde{Q}) \over \partial x^- \partial \tilde{x}^-}
= {\partial^2 \ln \gamma(x^-,Q,\tilde{x}^-,\tilde{Q}) \over \partial x^- \partial
\tilde{Q}^j_n}
$$
\eqn\ncuatro{
=
{\partial^2 \ln \gamma(x^-,Q,\tilde{x}^-,\tilde{Q}) \over \partial {Q}^j_n \partial
\tilde{x}^-}= {\partial^2 \ln \gamma(x^-,Q,\tilde{x}^-,\tilde{Q}) \over \partial {Q}^j_m 
\partial \tilde{Q}^k_n}= 0}
for every $j,k= 1, \dots , D-2$ and $m,n \in \IZ$  (we put $x^i \equiv Q^j_0,$ $p^j
\equiv P^j_0$). 

Conversely, let $\gamma$ satisfies Eq. (3.35). The general solution of (3.35) reads

\eqn\ncinco{
\gamma(x^-,Q,\tilde{x}^-,\tilde{Q}) = \Psi_1(x^-,Q) \Psi_2(\tilde{x}^-, \tilde{Q}).}
As the function $\rho_{_W}$ is assumed to be real we get from Eq. (3.34)

\eqn\nseis{
\gamma^*(x^-,Q,\tilde{x}^-,\tilde{Q}) = \gamma(\tilde{x}^-,\tilde{Q}, {x}^-,{Q}).}
Consequently, Eq. (3.36) has the form

\eqn\nsiete{
\gamma(x^-,Q,\tilde{x}^-,\tilde{Q}) = A \Psi_1(x^-,Q) \Psi_1^*(\tilde{x}^-,\tilde{Q}),}
where, by the assumption (3.28), $A$ is a positive real constant. Finally, defining
$\Psi:= \sqrt{A} \Psi_1(x^-,Q)$ one obtains

\eqn\nsiete{
\gamma(x^-,Q,\tilde{x}^-,\tilde{Q}) = \Psi(x^-,Q) \Psi^*(\tilde{x}^-, \tilde{Q}).}
Substituting $x^- \mapsto x^- + {\xi^- \over 2},$ $Q \mapsto Q + {\xi\over 2},$
$\tilde{x}^- \mapsto x^- - {\xi^- \over 2},$ $\tilde{Q} \mapsto Q - {\xi^- \over 2},$
multiplying both sides by $exp \big\{ - {i \over \hbar} (- \xi^- p^+ + \xi \cdot P)\big\}$
and integrating with respect to $ d({\xi^- \over 2 \pi \hbar}) d({ \xi \over 2 \pi \hbar})$
we get exactly the relation (3.21). This means that our function  $\rho_{_W}$ is the
Wigner function of the pure state $\Psi(x^-,Q)$. Thus we arrive at the following

\vskip 1truecm

\noindent
{\it Theorem 2.1}

A real function
$\rho_{_W}(x^-,Q,p^+,P)$ satisfying also the conditions (3.27) and (3.28) is  the
Wigner function of some pure state if and only if the function
$\gamma(x^-,Q,\tilde{x}, \tilde{Q})$
defined by (3.34) satisfies Eqs. (3.35) $\fund$. 

In terms of $(x^-,X,p^+, \Pi)$ variables the conditions (3.35) read

$$
{\partial^2 \ln \gamma[x^-,X,\tilde{x}^-,\tilde{X}] \over \partial x^- \partial \tilde{x}^-}
= {\partial \over \partial x^-} {\delta \ln \gamma[x^-,X,\tilde{x}^-,\tilde{X}] \over
\delta \tilde{X} }
$$
\eqn\nocho{
= {\partial \over \partial \tilde{x}^-}
{\delta \ln \gamma[x^-,X,\tilde{x}^-,\tilde{X}] \over \delta X}= 
{\delta^2 \ln
\gamma[x^-,X,\tilde{x}^-,\tilde{X}] \over \delta X  
\delta \tilde{X}}= 0.}

\vskip 1 truecm
\subsec{ Open Strings}

This is a simple matter to carry over the results of the preceding subsection to the
case of open string. The thing we must take in care is that the subindice $n$
standing at $Q^j_n$, $P^j_n$, $a^j_n$ and $a^{*j}_n$ takes the values $n=1, \dots , \infty$.
(We let also $n$ be zero in the formulas where $Q^j_0 \equiv x^j,$ $P^j_0 \equiv p^j$). 
Moreover, we should remember that now the oscillator frequencies  $\omega_n = n \in \IZ_+,$ and
not $2 |n|$, as before, and also that $\tilde{\alpha}_n,$ $\tilde{\alpha}_{-n}$ don't appear.
Thus in the present case one gets

$$ * = exp\bigg\{{i\hbar\over 2} \buildrel{\leftrightarrow}\over {\cal P}\bigg\}
$$

$$ 
= exp \bigg\{{i \hbar \over 2} \bigg[\bigg({{\buildrel{\leftarrow}\over {\partial}}\over
\partial p^{+}} {{\buildrel{\rightarrow}\over {\partial}}\over
\partial x^{-}} - {{\buildrel{\leftarrow}\over {\partial}}\over
\partial x^{-}} {{\buildrel{\rightarrow}\over {\partial}}\over
\partial p^{+}}\bigg) + \sum_{j=1}^{D-2} 
\bigg({{\buildrel{\leftarrow}\over {\partial}}\over
\partial x^{j}} {{\buildrel{\rightarrow}\over {\partial}}\over
\partial p^{j}} - {{\buildrel{\leftarrow}\over {\partial}}\over
\partial p^{j}} {{\buildrel{\rightarrow}\over {\partial}}\over
\partial x^{j}}\bigg) \bigg] \bigg \}$$

$$ exp \bigg\{ {1 \over 2}
\sum_{j=1}^{D-2} \sum_{n=1}^{\infty}
\bigg({{\buildrel{\leftarrow}\over {\partial}}\over
\partial a^{j}_n} {{\buildrel{\rightarrow}\over {\partial}}\over
\partial a^{*j}_n} - {{\buildrel{\leftarrow}\over {\partial}}\over
\partial a^{*j}_n} {{\buildrel{\rightarrow}\over {\partial}}\over
\partial a^{j}_n}\bigg)\bigg\}
$$

\eqn\nnueve{
= \dots exp \bigg\{ {\hbar \over 2}
\sum_{j=1}^{D-2} \sum_{n \not=0}
n {{\buildrel{\leftarrow}\over {\partial}}\over
\partial \alpha^{j}_n} {{\buildrel{\rightarrow}\over {\partial}}\over
\partial \alpha^{j}_{-n}} \bigg\}. }

Then the Wigner function of the ground state reads now

$$
\rho_{_{W0}} = C  exp \bigg \{ - 2  \sum_{j=1}^{D-2} \sum_{n=1}^{\infty}
a^j_n a^{*j}_n  \bigg\} \delta (p^1) \dots \delta(p^{D-2})
\delta(p^+)
$$

\eqn\cien{
=C  exp \bigg \{ - {1 \over \hbar} \sum_{j=1}^{D-2} \sum_{n=1}^{\infty}
{1\over n} \bigg( (P^j_n)^2 +  n^2 (Q^j_n)^2 \bigg) \bigg\} \delta (p^1) 
\dots \delta(p^{D-2}) \delta(p^+),}
with $C > 0$. (Compare with \dq\ Eq. (3.21)).

\vskip 2truecm
\newsec{Hamiltonian, Square Mass, Normal Ordering and Virasoro Algebra}

To proceed further with deformation quantization of bosonic string we must 
take into account that not only the Weyl ordering but also the normal ordering
should be implemented into this quantization. To this end we first consider the 
Casimir effect in string theory. Consider the real scalar field $\Phi(\tau,\sigma)$
on the cylindrical space time $\IR \times {\bf S}^1$. The circunference of ${\bf S}^1$
is $L$. The standard expansion of $\Phi(\tau,\sigma)$ satisfying the boundary conditions
$\Phi(\tau,\sigma) = \Phi(\tau,\sigma + nL)$ for all $n \in \IZ$ reads (compare with 
Eq. (2.12))

\eqn\cienuno{
 \Phi(\tau, \sigma) = x + {1 \over  L}p \tau
+ {1 \over \sqrt{L}} \sum_{n \not=0} \sqrt{{\hbar \over 2 \omega_n}}
 \bigg\{ a_n
exp\bigg( i({2 \pi n \over L} \sigma - \omega_n \tau) \bigg) +  a^{*}_n
exp\bigg( -i ({2 \pi n \over L}\sigma - \omega_n \tau) \bigg) \bigg\}, } 
where $\omega_n = {2 \pi |n| \over L}.$ The conjugate momentum $\Pi(\tau,\sigma) =
\dot{\Phi}(\tau,\sigma).$ Employing the deformation
quantization formalism one can show \dq\ that the expected value of the energy density
$\langle {T}_{00} \rangle (L)$ of the ground state is (the Casimir effect)

\eqn\ciendos{
\langle {T}_{00} \rangle (L)= - { \pi \hbar \over 6 L^2}.}
(In terms of the usual quantum field theory see [32-34].)

Consequently, for the total energy $E_0(L)$ of the ground state one gets

\eqn\cientres{
E_0(L)= L \cdot \langle {T}_{00} \rangle (L) = - {\pi \hbar \over 6 L}.}

Consider now the real scalar field $\Phi(\tau,\sigma)$ on $\IR \times [0,L]$ but with the
boundary conditions ${\partial \Phi(\tau,0) \over \partial \sigma} = 0 = {\partial
\Phi(\tau,L)  \over \partial \sigma}$ for all $\tau \in \IR$. It is a simple matter to show
that now the expansion of $\Phi(\tau,\sigma)$ takes the following form

\eqn\ciencuatro{
 \Phi(\tau, \sigma) = x + {1 \over  L}p \tau
+ {1 \over \sqrt{2L}} \sum_{n \not=0} \sqrt{{\hbar \over 2 \omega_n}}
 \bigg\{ a_n
exp\bigg( i({ \pi n \over L} \sigma - \omega_n \tau) \bigg) +  a^{*}_n
exp\bigg( -i ({ \pi n \over L}\sigma - \omega_n \tau) \bigg) \bigg\}, } 
where $\omega_n = { \pi |n| \over L}$ and $a_n = a_{-n}$.
Comparing (4.1) with (4.4) one quickly arrives at the conclusion that
the oscillating part in (4.4) is {\it mutatis mutandi} the same as in (4.1) if in (4.1) $L$ is
changed to $2L$. Hence it follows that in the present case the Casimir effect can be obtained
from (4.2) by inserting $2L$ instead of $L$. Thus we have now

\eqn\ciencinco{
\langle {T}_{00} \rangle (L) = - { \pi \hbar \over 24 L^2}}
and for the total energy of the ground state

\eqn\cienseis{
E_0(L) = L \cdot \langle {T}_{00} \rangle (L) = - {\pi \hbar \over 24 L}.}
We use the above results to the deformation quantization of bosonic strings.

\vskip 1truecm
\subsec{Closed Strings}

In this case one can consider $X^j(\tau, \sigma)$, $j =1, \dots , D-2,$
to be $D-2$ real scalar massless fields on the cylindrical worldsheet
$\IR \times {\bf S}^1$ with $L = \pi.$ Therefore, by Eq. (4.3), the vacuum
energy $E_0$ reads now

\eqn\ciensiete{
E_0= - {\hbar (D-2) \over 6} = : - 4a.}
To obtain this $E_0$ from the eigenvalue equation we put $\hat{\cal N} ' H$

$$
\hat{\cal N} ' := exp \bigg\{ \sum_{j=1}^{D-2} \sum_{n \not= 0}( - {1\over 2}
+ \gamma_n ) {\partial^2 \over \partial a^j_n \partial a^{*j}_n} \bigg\}
$$

\eqn\cienocho{
= exp \bigg\{ \hbar \sum_{j=1}^{D-2} \sum_{n=1}^{\infty} n \bigg(
(- {1\over 2} + \gamma_n ){ \partial^2 \over \partial \alpha^j_n \partial \alpha^j_{-n}}
+ (- {1 \over 2} + \gamma_{-n}){ \partial^2 \over \partial \tilde{\alpha}^j_n
\partial \tilde{\alpha}^j_{-n}} \bigg) \bigg\}}
instead of the Hamiltonian $H$ given by Eq. (2.34). Then

\eqn\ciennueve{
\hat{\cal N} ' H * \rho_{_{W0}} = -4 a \rho_{_{W0}} }
if 

\eqn\ciendiez{
\sum_{n \not= 0} |n| \gamma_n = - {D-2 \over 12}.}
From the previous work \dq\ we know that the Weyl image of $\hat{\cal N} ' H$ reads

\eqn\cienonce{
W(\hat{\cal N} ' H ) = \ \ : W(H): \ - 4a, }
where $:W(H):$ is the normal ordered operator $W(H)$ and it can be written as follows

$$
:W(H): \ = W(\hat{\cal N} H )
$$

\eqn\ciendoce{
\hat{\cal N}  := exp \bigg\{ -{1 \over 2}\sum_{j=1}^{D-2} \sum_{n \not= 0}
 {\partial^2 \over \partial a^j_n \partial a^{*j}_n} \bigg\}
= exp \bigg\{-{\hbar \over 2} \sum_{j=1}^{D-2} \sum_{n=1}^{\infty} n \bigg(
{ \partial^2 \over \partial \alpha^j_n \partial \alpha^j_{-n}}
+ { \partial^2 \over \partial \tilde{\alpha}^j_n
\partial \tilde{\alpha}^j_{-n}} \bigg) \bigg\}.}

Analogously for the square mass given by Eq. (2.32) we get

\eqn\cientrece{
\hat{\cal N} ' M^2 * \rho_{_{W0}} = - 8 \pi T a \rho_{_{W0}}.}
One quickly finds that

\eqn\cientreceuno{ 
W(\hat{\cal N} ' M^2) = W(\hat{\cal N}M^2) - 8 \pi Ta =
:W(M^2): \ - 8 \pi Ta.}

Given the {\it normal ordering operator} $\hat{\cal N}$ and the {\it generalized normal ordering
operator} $\hat{\cal N} '$ one can define new star products which are cohomologically equivalent
to the Moyal $*$-product (see Eqs. (3.13) or (3.18)). These star products will be denoted by $*_{\cal N}$
and
$*_{{\cal N} '}$ respectively and they are given by

$$
F_1 *_{\cal N}  F_2 = \hat{\cal N}^{-1} \bigg( \hat{\cal N}F_1 * \hat{\cal N} F_2 \bigg)
$$

\eqn\ciencatorce{
F_1 *_{{\cal N} '}  F_2 = \hat{\cal N} '^{-1} \bigg( \hat{\cal N} 'F_1 * \hat{\cal N} 'F_2 \bigg).}
Consequently the eigenvalue equations for the Hamiltonian or the square mass read (compare with
Eq. (4.9) or (4.13))

$$
H  *_{{\cal N} '} \rho_{_W}^{{\cal N} '} = E \rho_{_W}^{{\cal N} '} \Longrightarrow
H  *_{{\cal N} '} \rho_{_{W0}}^{{\cal N} '} = -4a\rho_{_{W0}}^{{\cal N} '}
$$

\eqn\cienquince{
M^2  *_{{\cal N} '} \rho_{_W}^{{\cal N} '} = \mu^2 \rho_{_W}^{{\cal N} '} \Longrightarrow
M^2  *_{{\cal N} '} \rho_{_{W0}}^{{\cal N} '} = -8 \pi Ta \rho_{_{W0}}^{{\cal N} '}}
where $\rho_{_{W}}^{{\cal N} '} := \hat{\cal N}^{-1} \rho_{_W}.$

It is an easy matter to show that

$$ 
\alpha_{-n} *_{{\cal N}} \alpha_n = \alpha_{-n} \alpha_n, \ \ \ \ \ 
\alpha_n *_{{\cal N}} \alpha_{-n} = \alpha_n \alpha_{-n} + \hbar n
$$

\eqn\ciendseis{
\tilde{\alpha}_{-n} *_{{\cal N}} \tilde{\alpha}_n = \tilde{\alpha}_{-n} \tilde{\alpha}_n, \ \ \ \ \ 
\tilde{\alpha}_n *_{{\cal N}} \tilde{\alpha}_{-n} = \tilde{\alpha}_n \tilde{\alpha}_{-n} + \hbar n}
for all $n \in \IZ_+$. All other products are the usual products.

Taking into account Eq. (4.17) one gets

$$
\{\alpha^-_m, \alpha^-_n \}^{({\cal N})} = -i {2 \sqrt{\pi T} \over p^+}(m-n) \hat{\cal N} \alpha^-_{m+n}
-i \hbar {4 \pi T \over (p^+)^2} {D-2 \over 12} m(m^2 -1) \delta_{m+n,0}
$$
\eqn\ciendsiete{
\{\tilde{\alpha}^-_m, \tilde{\alpha}^-_n \}^{({\cal N})} = -i {2 \sqrt{\pi T} \over p^+}(m-n) \hat{\cal N}
\tilde{\alpha}^-_{m+n}
-i \hbar {4 \pi T \over (p^+)^2} {D-2 \over 12} m(m^2 -1) \delta_{m+n,0},}
where $\{ \alpha^-_m, \alpha^-_n \}^{({\cal N})}: = {1 \over i \hbar} \big( \alpha^-_m *_{{\cal N}} \alpha^-_n
- \alpha^-_n *_{{\cal N}} \alpha^-_m \big)$ etc.
Thus we arrive at the {\it Virasoro algebra with a central extension}.

\noindent
$\big[$ {\it Remark}: Calculations in Eq. (4.18) are rather formal. To perform them one must always put
$\alpha_n$ on the right to $\alpha_{-m}$ $m,n \in \IZ_+$, $m \not= n$, and the same for $\tilde{\alpha}_n$ and
$\tilde{\alpha}_{-m}$. See the analogous calculations in terms of operator language
\theisen. $\big]$

\vskip 1truecm
\subsec{ Open Strings}
Here we can find the energy of the vacuum state $E_0$ by substituting $L= \pi$ into (4.6) and taking into
account that we deal with $D-2$ scalar fields. Hence 

\eqn\ciendocho{
E_0 = - { \hbar (D-2) \over 24} = -a. }
Now the normal ordering operator $\hat{\cal N}$ and the generalized normal ordering operator $\hat{\cal N} '$
read

$$
\hat{\cal N} = exp \bigg\{ -{1 \over 2}\sum_{j=1}^{D-2} \sum_{n=1}^{\infty}
{\partial^2 \over \partial a^j_n \partial a^{*j}_n} \bigg\}
= exp \bigg\{ -{\hbar \over 2}\sum_{j=1}^{D-2} \sum_{n=1}^{\infty}
n {\partial^2 \over \partial \alpha^j_n \partial \alpha^{j}_{-n}} \bigg\},
$$

\eqn\ciendnueve{
\hat{\cal N} ' = 
= exp \bigg\{ \sum_{j=1}^{D-2} \sum_{n=1}^{\infty}  \bigg(
(- {1\over 2} + \beta_n){ \partial^2 \over \partial a^j_n \partial a^{*j}_{n}}\bigg) \bigg\}
= exp \bigg\{ \hbar \sum_{j=1}^{D-2} \sum_{n=1}^{\infty}  n \bigg(
(- {1\over 2} + \beta_n){ \partial^2 \over \partial \alpha^j_n \partial \alpha^{j}_{-n}}\bigg) \bigg\}
}
where

\eqn\cienveinte{
 \sum_{n=1}^{\infty} n \beta_n = - {D-2 \over 24}.}
Then

$$
H  *_{{\cal N} '} \rho_{_W}^{{\cal N} '} = E \rho_{_W}^{{\cal N} '} \Longrightarrow
H  *_{{\cal N} '} \rho_{_{W0}}^{{\cal N} '} = -a\rho_{_{W0}}^{{\cal N} '}
$$

\eqn\cienvuno{
M^2  *_{{\cal N} '} \rho_{_W}^{{\cal N} '} = \mu^2 \rho_{_W}^{{\cal N} '} \Longrightarrow
M^2  *_{{\cal N} '} \rho_{_{W0}}^{{\cal N} '} = -2 \pi Ta \rho_{_{W0}}^{{\cal N} '}.}
Thus, as before, the ground state is the tachyonic one. 

Finally, the {\it Virasoro algebra with a central extension} reads now

\eqn\cienvdos{
\{\alpha^-_m, \alpha^-_n \}^{({\cal N})} = -i { \sqrt{\pi T} \over p^+}(m-n) \hat{\cal N} \alpha^-_{m+n}
-i \hbar {\pi T \over (p^+)^2} {D-2 \over 12} m(m^2 -1) \delta_{m+n,0}.}

\vskip 2truecm

\newsec{ Some Simple Example: The Wightman Functions}

Here we are going to present a simple example of calculations within the deformation 
quantization formalism. Namely, we find the Wightman (Green) functions 
$\langle X^j(\tau, \sigma) * X^k(\tau ', \sigma ') \rangle$. By the definition
(see(3.23))

\eqn\cienvtres{
\langle X^j(\tau,\sigma) * X^k(\tau ',\sigma ') \rangle
= { \int d x^- d p^+ dQ dP
\rho_{_{W0}}(x^-,Q,p^+,P)  X^j(\tau,\sigma) * X^k(\tau ',\sigma ')
\over \int d x^- d p^+ dQ dP \rho_{_{W0}}(x^-,Q,p^+,P)}.}
where $\rho_{_{W0}}$ is given by Eq. (3.32) (closed string) or Eq. (3.42) (open string)
and $X^j(\tau,\sigma)$ is given by Eq. (2.41) (closed string) or by Eq. (2.56) (open string).

\vskip 1truecm
\subsec{Closed Strings}
Employing Eqs. (2.42) and (3.18) and perfoming simple integrations (Gaussian integrals)
one gets

\eqn\cienvcuatro{
\langle Q^j_m(\tau) * Q^k_n(\tau ') \rangle = \delta_{jk} \delta_{mn}
{\hbar \over 4 |m|} exp \bigg( -2i |m|(\tau - \tau ') \bigg), \ \ \ m,n \not=0.}
Hence, differentiating (5.2) with respect to $\tau$ or/and $\tau '$ we obtain

$$
\langle Q^j_m(\tau) * P^k_n(\tau ') \rangle = \delta_{jk} \delta_{mn} 
{i \hbar \over 2} exp \bigg( -2i |m|(\tau - \tau ') \bigg) = - \langle P^j_m(\tau) *
Q^k_n(\tau ') \rangle
$$

\eqn\cienvcinco{
\langle P^j_m(\tau) * P^k_n(\tau ') \rangle = \delta_{jk} \delta_{mn} 
\hbar |m| exp \bigg( -2i |m|(\tau - \tau ') \bigg), \ \ \ \  m,n \not= 0.}
Then we have also

$$
\langle x^j * p^k \rangle = \delta_{jk} {i \hbar \over 2} = - \langle p^j * x^k \rangle
$$

\eqn\cienvseis{
\langle p^j * p^k \rangle = 0,  \ \ \ \ \ \ \  \langle x^j * x^k
\rangle = \delta_{jk} \langle (x^j)^2 \rangle.}
Using Eqs. (5.2), (5.3) and (5.4) one easily finds

\eqn\cienvsiete{
\langle X^j(\tau,\sigma) * X^k(\tau ',\sigma ') \rangle = \delta_{jk}
\bigg\{ \langle x^j x^k\rangle + {i \hbar \over 2 \pi T} (\tau '- \tau)
+ {\hbar \over 4 \pi T} \sum_{n \not= 0}
{exp\bigg( 2i [n (\sigma - \sigma') - |n|(\tau - \tau ')]\bigg) \over
|n|} \bigg\}.}
Performing the summations in Eq. (5.5) and removing the part independent of the 
coordinates $(\tau, \sigma, \tau ', \sigma ')$ we have 

\eqn\cienvocho{
\langle X^j(\tau,\sigma) * X^k(\tau ',\sigma ') \rangle \sim \delta_{jk}
 \big(- {\hbar \over 4 \pi T}\big) \bigg\{ \ln |\sin \big[ \tau ' - \sigma ' -(\tau -\sigma)
\big] | +  \ln |\sin \big[ \tau ' + \sigma ' -(\tau +\sigma)
\big] | \bigg\}.}
(Compare with [1,2,4,33]).

\vskip 1truecm
\subsec{Open Strings}

For open strings we get

$$
\langle Q^j_m(\tau) * Q^k_n(\tau ') \rangle = \delta_{jk} \delta_{mn}
{\hbar \over 2 |m|} exp \bigg( -i |m|(\tau - \tau ') \bigg),
$$
$$
\langle Q^j_m(\tau) * P^k_n(\tau ') \rangle = \delta_{jk} \delta_{mn} 
{i \hbar \over 2} exp \bigg( -i |m|(\tau - \tau ') \bigg) = - \langle P^j_m(\tau) *
Q^k_n(\tau ') \rangle,
$$

\eqn\cienvnueve{
\langle P^j_m(\tau) * P^k_n(\tau ') \rangle = \delta_{jk} \delta_{mn} 
{\hbar |m| \over 2} exp \bigg( -i |m|(\tau - \tau ') \bigg), \ \ \ \  m,n \not= 0.}
Then

\eqn\cientreinta{
\langle X^j(\tau,\sigma) * X^k(\tau ',\sigma ') \rangle = \delta_{jk}
\bigg\{ \langle x^j x^k  \rangle + {i \hbar \over 2 \pi T} (\tau '- \tau)
+ {\hbar \over \pi T} \sum_{n= 1}^{\infty}
{exp\big( -i n (\tau - \tau ')\big) \cos (n \sigma) \cos( n \sigma ') 
 \over
n} \bigg\}.}
Performing the summations in Eq. (5.8) and removing the part independent of the 
coordinates we get 

$$
\langle X^j(\tau,\sigma) * X^k(\tau ',\sigma ') \rangle \sim \delta_{jk}
 \big(- {\hbar \over 4 \pi T}\big) \bigg\{ \ln |\sin \bigg( {\tau ' - \sigma ' -(\tau
-\sigma)\over 2}\bigg) | +  \ln |\sin \bigg( {\tau ' - \sigma ' -(\tau +\sigma)\over
2}\bigg) |
$$

\eqn\cientuno{ 
+\ln |\sin \bigg( {\tau ' + \sigma ' -(\tau - \sigma)\over 2} \bigg) | 
+\ln |\sin \bigg( {\tau ' + \sigma ' -(\tau +\sigma)\over 2} \bigg) | \bigg\}.}

\vskip 2truecm

\newsec{ Final Remarks}

In this paper we have investigated bosonic string theory within the
deformation quantization formalism. As it is seen deformation quantization 
provides us with a tool which enables one to describe quantum bosonic string 
in terms of deformed Poisson-Lie algebra.

The most intriguing problem is to obtain the similar result for the case of 
superstring theory. Some works on defomation quantization of supersymmetric field
theory (see e.g. [35,36,37]) seem to be crucial in searching for a solution for the
problem. Another interesting possibility is the extension of the matters considerd here, to
the case of interacting strings and interacting superstrings. We are going to consider these
problems in the near future.

\vskip 2truecm

\centerline{\bf Acknowledgements}

This paper was partially supported by CINVESTAV and CONACyT (M\'exico) and by
KBN (Poland).

One of us (M.P.) is indebted to the staff of Departmento de F\'{\i}sica at
CINVESTAV (M\'exico) for warm hospitality.


\listrefs

\end